\documentclass[Physsubmission, Phys]{SciPost}

\hypersetup{
    colorlinks,
    linkcolor={red!50!black},
    citecolor={blue!50!black},
    urlcolor={blue!80!black}
}

\usepackage[bitstream-charter]{mathdesign}
\usepackage{amsmath}

\DeclareSymbolFont{usualmathcal}{OMS}{cmsy}{m}{n}
\DeclareSymbolFontAlphabet{\mathcal}{usualmathcal}

\begin{document}

\begin{center}{\Large \textbf{
Collective dynamics of heavy ion collisions in ATLAS
}}\end{center}

\begin{center}
Helena Santos\textsuperscript{*}, on behalf of the ATLAS Collaboration
\end{center}

\begin{center}
{\bf } Laborat\'orio de Instrumenta\c{c}\~ao e F\'isica Experimental de Part\'iculas, LIP\\ Av. Prof. Gama Pinto 2, Lisbon, Portugal\\
* helena@lip.pt
\end{center}

\begin{center}
\today
\end{center}


\definecolor{palegray}{gray}{0.95}
\begin{center}
\colorbox{palegray}{
  \begin{tabular}{rr}
  \begin{minipage}{0.1\textwidth}
    \includegraphics[width=30mm]{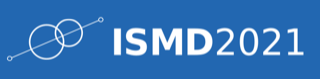}
  \end{minipage}
  &
  \begin{minipage}{0.75\textwidth}
    \begin{center}
    {\it 50th International Symposium on Multiparticle Dynamics}\\ {\it (ISMD2021)}\\
    {\it 12-16 July 2021} \\
    \doi{10.21468/SciPostPhysProc.?}\\
    \scriptsize{Copyright 2021 CERN for the benefit of the ATLAS Collaboration. CC-BY-4.0 license.}
    \end{center}
  \end{minipage}
\end{tabular}
}
\end{center}

\section*{Abstract}
{\bf
The latest measurements of collective behaviour in a variety of collision systems with the ATLAS detector at the LHC, including \mbox{$pp$} collisions at 13 TeV, \mbox{Xe+Xe} collisions at
5.44 TeV, and $\mbox{Pb+Pb}$ collisions at 5.02 TeV, are presented. They include  $v_n$-[\mbox{$p_{\mathrm{T}}$}] correlations, which carry important information about the initial-state geometry of the
quark-gluon plasma and can shed light on any quadrupole deformation in the Xe nucleus, and measurements of flow decorrelations differential in rapidity, which probe the longitudinal structure of
the colliding system. These measurements furthermore provide stringent tests of the theoretical understanding of the initial state in heavy ion collisions.\\}
\section{Introduction}
\label{sec:intro}
Heavy-ion collisions at the LHC produce the quark-gluon plasma (QGP) whose space-time evolution is described by hydrodynamics\cite{hydro1}. Owing to strong event-by-event density fluctuations in the initial state, the space-time evolution of the QGP also fluctuates event by event. These fluctuations lead to correlations of particle multiplicity in momentum space in both the transverse and longitudinal directions with respect to the collision axis. Studies of particle correlations in the transverse plane reveal strong harmonic modulation of the particle densities in the azimuthal angle, $\frac{dN}{d\phi} \equiv \frac{N_{0}}{2\pi}(1+2\sum_{n=1}^{\infty}v_{n}{\rm cos}(n(\phi-\Phi_{n})))$, where $v_n$ and $\Phi_n$ represent the magnitude and event plane angle\footnote{The event plane is defined by the beam direction and by the direction of the impact parameter $b$.} of the $n^\mathrm{th}$-order azimuthal flow vector $V_n = v_ne^{in\Phi_n}$.\\
The \textsc{Glauber} Monte Carlo model\cite{Glauber} is used to obtain a correspondence between the total transverse energy deposited in the ATLAS\cite{ATLAS} forward calorimeter (FCal) and the sampling fraction of the total inelastic \mbox{A+A} cross-section, allowing the setting of the centrality percentiles.
\section{$v_n$-[\mbox{$p_{\mathrm{T}}$}] correlations}
\label{sec:vnpt}
\begin{figure}[h]
\centering
\includegraphics[width=0.7\textwidth]{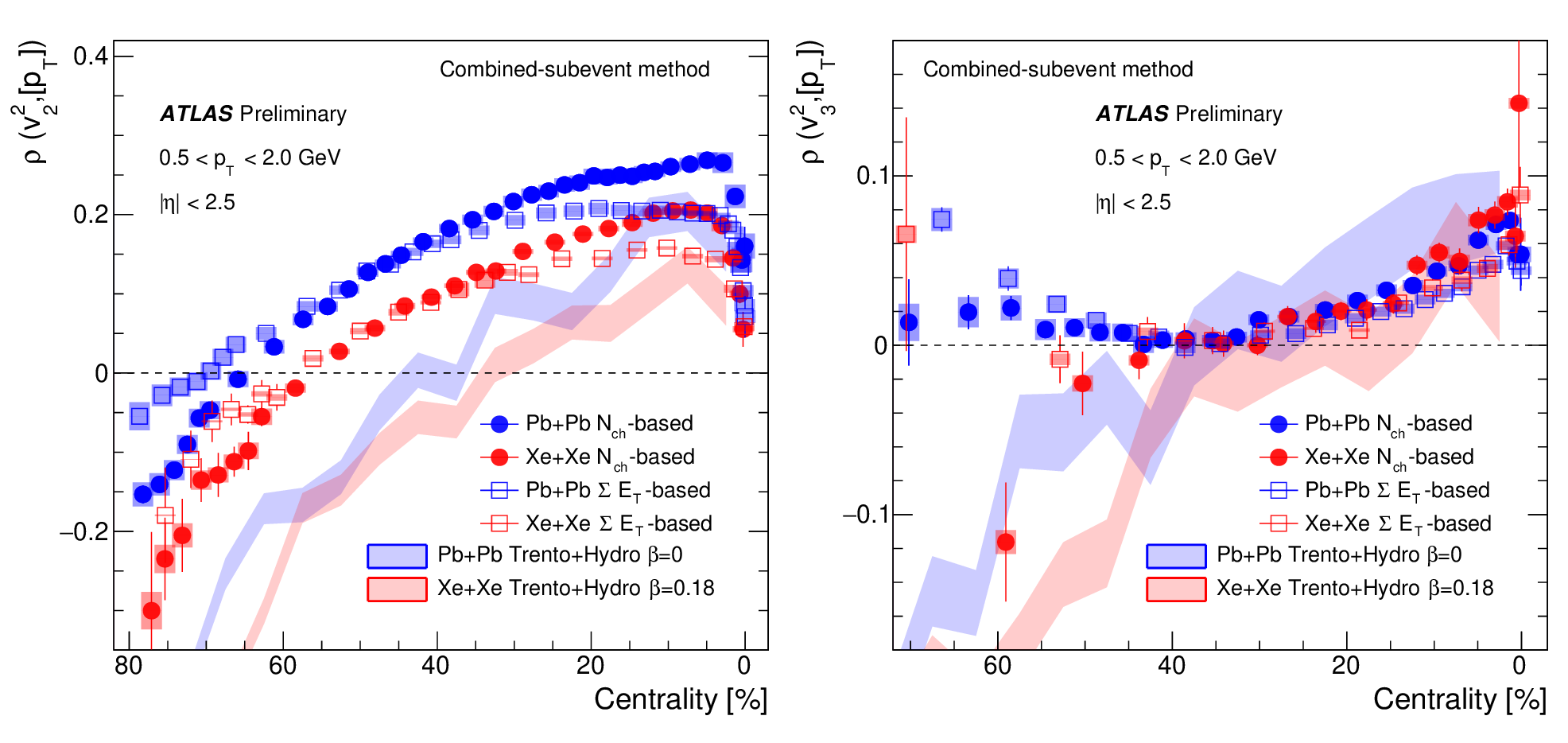}\\
\includegraphics[width=0.7\textwidth]{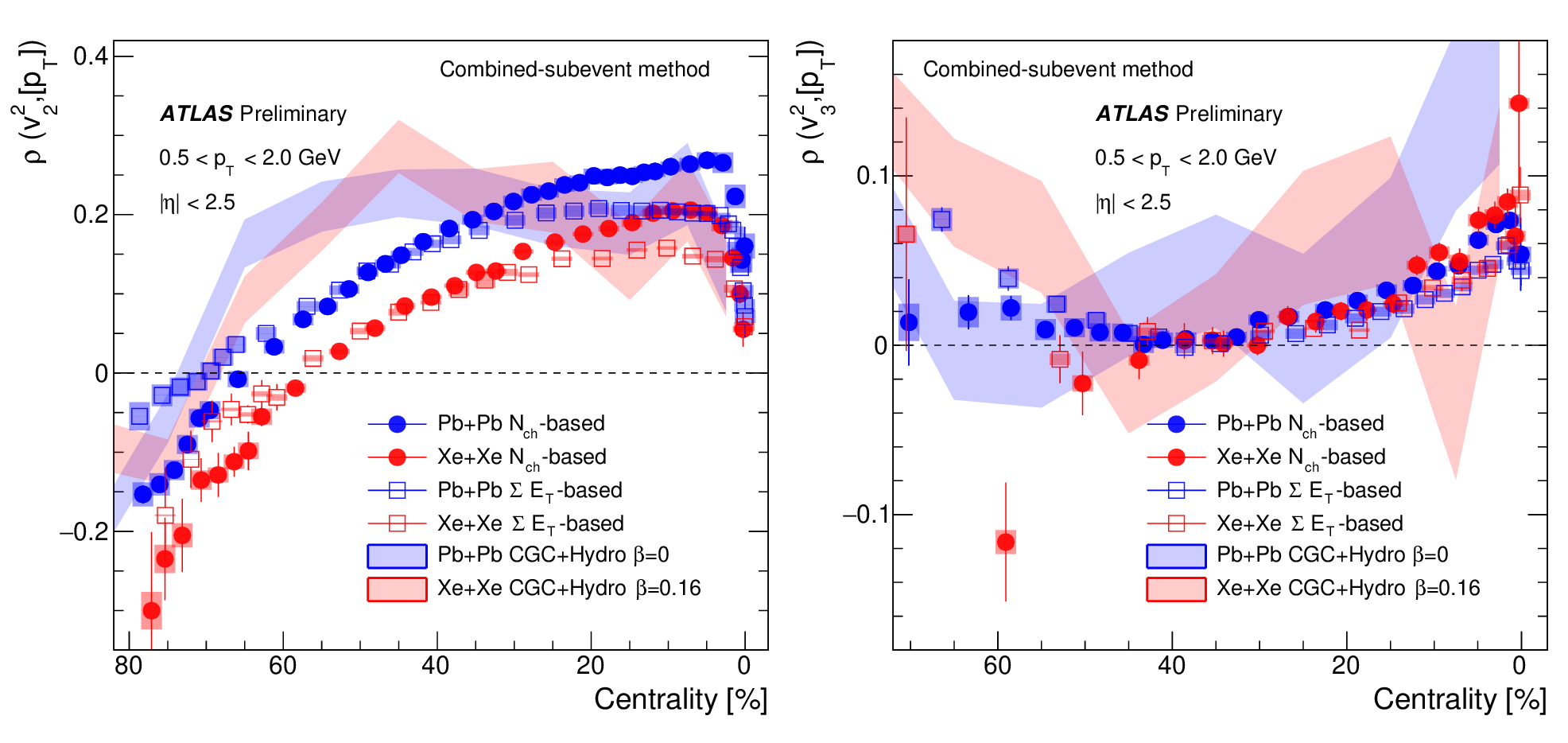}
\caption{$\rho(v_2^2,[p_{\mathrm{T}}])$ (left) and $\rho(v_3^2,[p_{\mathrm{T}}])$ (right) as a function of the collisions centrality in \mbox{Pb+Pb} (blue symbols) and \mbox{Xe+Xe} (red symbols) collisions obtained using $N^{\mathrm{ch}}_{\mathrm{rec}}$-based (closed symbols) and $\Sigma E_{\mathrm{T}}$-based (open symbols) event averaging procedure for charged particles in 0.5$<p_{\mathrm{T}}<$2 GeV\cite{ATLAS1}. They are compared with a hydrodynamical model calculation based on Trento initial condition\cite{trento2} (top) and with a calculation based on a three-dimensional initial condition dynamically generated from gluon saturation models\cite{gluon2} (bottom). The error bars and boxes on the data points regard statistical and systematic uncertainties, respectively. The width of the bands represent the statistical uncertainties of the models.}
\label{Figure:cor}
\end{figure}

The goal of this measurement is to understand the system size dependence, and in particular the role of the Xe nucleus deformation, through comparisons between \mbox{Pb+Pb} and \mbox{Xe+Xe} collisions results. Model calculations show that $V_n$ are  proportional to the eccentricities $\mathcal{E}_n$ for $n$ = 2, 3, and 4 in central collisions\cite{vnen}. Correlated fluctuations between the $\mathcal{E}_n$ and the system size in the initial state are expected to generate dynamical correlations between $v_n$ and [\mbox{$p_{\mathrm{T}}$}] in the final state. The correlator $\rho(v_n^2,[p_{\mathrm{T}}]) = cov(v_n^2,[p_{\mathrm{T}}]) /(\sqrt(var(v_n^2))\sqrt(c_k))$, where $[p_{\mathrm{T}}]$ is the average transverse momentum of particles in each event and $c_k$ is the variance of $[p_{\mathrm{T}}]$, is shown in Figure~\ref{Figure:cor}. A smaller magnitude of $\rho(v_2^2,[p_{\mathrm{T}}]) $ in \mbox{Xe+Xe} collisions is observed in all centrality range and there is a significant difference in the $N^{\mathrm{ch}}_{\mathrm{rec}}$-based and $\Sigma E_{\mathrm{T}}$-based (number of reconstructed charged particles and sum of the transverse energy depositions in the FCal, respectively) centrality binning. These systematic differences are much smaller in $\rho(v_3^2,[p_{\mathrm{T}}])$. The models \textsc{Trento+Hydro}\cite{trento2}  and \textsc{CGC+Hydro}\cite{gluon2} do not capture the trends in the data, either qualitatively or quantitatively. Although the correlations between $v_n$ and [\mbox{$p_{\mathrm{T}}$}] are sensitive to the nuclear deformations in the initial state, centrality fluctuations need to be taken into account in the understanding of the nuclear deformation effects on the \mbox{Xe+Xe} results\cite{ATLAS1}.
\section{Longitudinal decorrelation dynamics}
\label{sec:decor}
\begin{figure}[h]
  \centerline{
    \includegraphics[width=0.42\textwidth, angle=0]{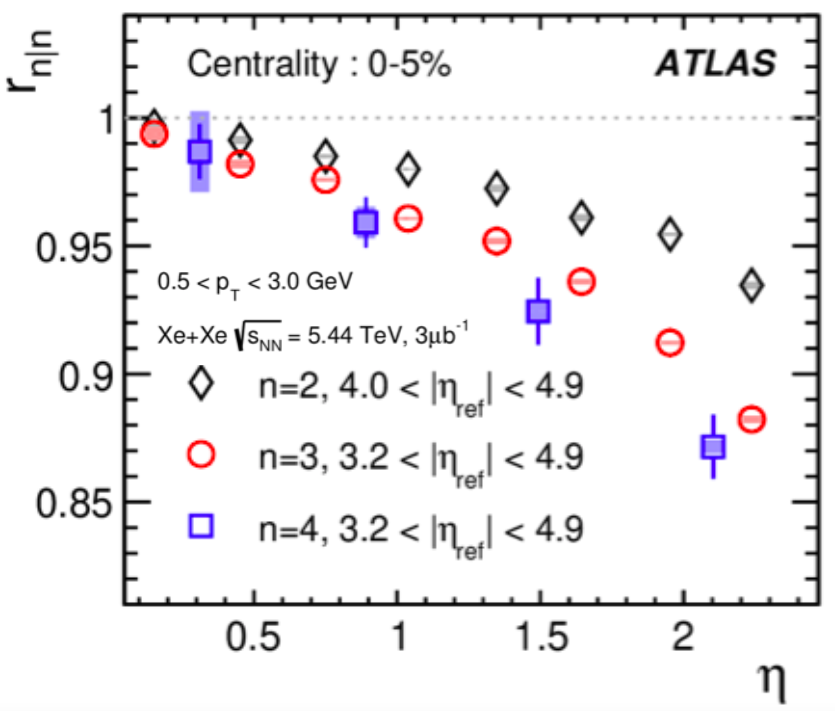}\\
    \includegraphics[width=0.47\textwidth, angle=0]{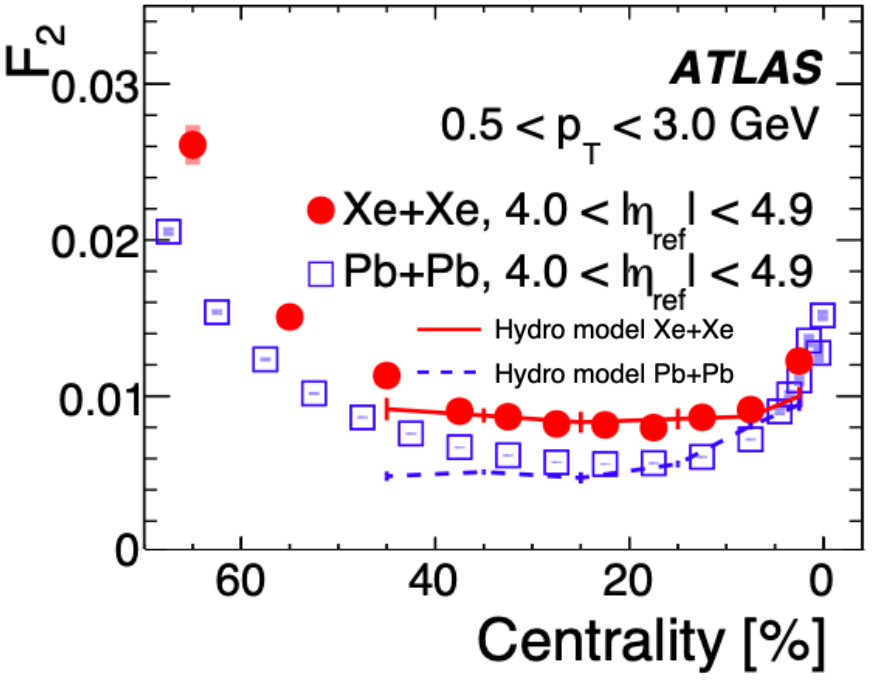}\\
  }
\caption{Left panel: $r_{2|2}$, $r_{3|3}$ and $r_{4|4}$ as a function of $\eta$ in \mbox{Xe+Xe} collisions for six centrality intervals. The $|\eta_{\mathrm{ref}}|$ stands between 4.0$<|\eta_{\mathrm{ref}}|<$4.9 for $r_{2|2}$, and 3.2$<|\eta_{\mathrm{ref}}|<$4.9 for $r_{3|3}$ and $r_{4|4}$. The error bars and boxes regard statistical and systematic uncertainties, respectively.  Right panel: The $F_2$ compared between Xe+Xe and Pb+Pb collisions as a function of centrality. The error bars and shaded boxes on the data represent statistical and systematic uncertainties, respectively\cite{ATLAS2} . The results from a hydrodynamic model are shown as solid lines (\mbox{Xe+Xe}) and dashed lines (\mbox{Pb+Pb}) with the vertical error bars representing statistical uncertainty of the predictions\cite{long2} . }
\label{Figure:decor}
\end{figure}

The longitudinal flow decorrelations are studied using the product of the particles weighted flow vectors $q_n(\eta)=\Sigma_j w_je^{in\phi_j}/(\Sigma_jw_j)$ in the Inner Detector\cite{ATLAS} and $q_n(\eta_{\mathrm{ref}})$ in the FCal, averaged over events in a given centrality interval. The $r_{n|n}(\eta)$ correlator, which quantifies the decorrelation between $\eta$ and -$\eta$, and the slope parameter $F_2$, which is obtained via a simple linear regression of the $r_{n|n}(\eta)$ data, are shown in Figure~\ref{Figure:decor} for \mbox{Xe+Xe} collisions. The  $r_{2|2}$ > $r_{3|3}$ > $r_{4|4}$ decrease linearly with $\eta$ and $F_2$ shows a strong centrality dependence, being smallest in the 20–30\% centrality interval and larger towards more-central and more-peripheral collisions. This strong centrality dependence is due to the average elliptic geometry in mid-central collisions, which dominates $v_2$ and makes this coefficient less sensitive to decorrelations, while the fluctuation-driven collision geometries dominate in central and peripheral collisions\cite{ATLAS2}.
\vspace{-0.5cm}
\paragraph{Funding information}
\scriptsize{The author acknowledges the financial support of Funda\c{c}\~ao para a Ci\^encia e a Tecnologia (FCT) through FCT Researcher contracts CEECIND/03346/2017 and CERN/FIS-PAR/0002/2019.}


\begin{thebibliography}{99}
\bibitem{hydro1} 
C. Gale, S. Jeon and B. Schenke, Int. J. Mod. Phys. A  {\bf 28}, 1340011 (2013). \doi{10.1142/S0217751X13400113}.
\bibitem{Glauber} 
M.L. Miller et al., Ann. R. Nucl. Part. Sci., {\bf57} 205 (2007). \doi{10.1146/annurev.nucl.57.090506.123020}.
\bibitem{ATLAS}
ATLAS Collaboration, JINST {\bf3}, S08003 (2008). \doi{10.1088/1748-0221/3/08/S08003}.
\bibitem{vnen}
F. G. Gardim, F. Grassi, M. Luzum and J.-Y. Ollitrault, Phys. Rev. C {\bf85}, 024908 (2012). \doi{10.1103/PhysRevC.85.024908}
\bibitem{ATLAS1}
ATLAS Collaboration, https://cds.cern.ch/record/2748818/files/ATLAS-CONF-2021-001.pdf.
\bibitem{trento2} 
G. Giacalone,  https://arxiv.org/abs/2101.00168 [nucl-th].
\bibitem{gluon2} 
G. Giacalone, B. Schenke and C. Shen, Phys. Rev. Lett. {\bf 125}, 192301 (2020). \doi{10.1103/PhysRevLett.125.192301}.
\bibitem{ATLAS2}
ATLAS Collaboration, Phys. Rev. Lett. {\bf 126}, 12230 (2021). \doi{10.1103/PhysRevLett.126.122301}.
\bibitem{long2} 
X.-Y. Wu, L.-G. Pang, G.-Y. Qin, and X.-N. Wang,  Phys. Rev. C {\bf 98}, 024913 (2018). \doi{10.1103/PhysRevC.98.024913}.
\end{thebibliography}
\end{document}